\documentclass[aps,pre,twocolumn,superscriptaddress,10pt]{revtex4}
\usepackage[dvips]{graphics}
\usepackage{graphicx}
\usepackage{amsfonts}
\usepackage{color}
\usepackage{amssymb}
\usepackage{amsmath}

\begin{document}

\title{ Controlling the transverse instability of dark solitons \\
and nucleation of vortices by a potential barrier }
\author{Manjun Ma}
\affiliation{College of Science, China Jiliang University, Hangzhou, Zhejiang, 310018,
P.R. China}
\author{R. Carretero-Gonz\'{a}lez\thanks{%
Corresponding author. On sabbatical leave from: Nonlinear Dynamical System
Group (\texttt{http://nlds.sdsu.edu}), Computational Science Research
Center, and Department of Mathematics and Statistics, San Diego State
University, San Diego, California 92182-7720, USA}}
\affiliation{Nonlinear Physics Group, Departamento de F\'{\i}sica Aplicada I, Universidad
de Sevilla, Avda.~Reina Mercedes s/n., 41012 Sevilla, Spain}
\author{P. G. Kevrekidis}
\affiliation{Department of Mathematics and Statistics, University of Massachusetts,
Amherst MA 01003-4515}
\author{D. J. Frantzeskakis}
\affiliation{Department of Physics, University of Athens, Panepistimiopolis, Zografos,
Athens 15784, Greece}
\author{B. A. Malomed}
\affiliation{Department of Physical Electronics, School of Electrical Engineering,
Faculty of Engineering, Tel Aviv University, Tel Aviv 69978, Israel}
\date{\today }

\begin{abstract}
We study possibilities to suppress the transverse modulational instability
(MI) of dark-soliton stripes in two-dimensional (2D) Bose-Einstein
condensates (BECs) and self-defocusing bulk optical waveguides by means of
quasi-1D structures. Adding an external repulsive barrier potential (which
can be induced in BEC by a laser sheet, or by an embedded plate in optics),
we demonstrate that it is possible to reduce the MI wavenumber band, and
even render the dark-soliton stripe completely stable. Using this method, we
demonstrate the control of the number of vortex pairs nucleated by each
spatial period of the modulational perturbation. By means of the
perturbation theory, we predict the number of the nucleated vortices per
unit length. The analytical results are corroborated by the numerical
computation of eigenmodes of small perturbations, as well as by direct
simulations of the underlying Gross-Pitaevskii/nonlinear Schr\"{o}dinger
equation.
\end{abstract}

\maketitle

\section{Introduction}

Since the experimental creation of atomic Bose-Einstein condensates
(BECs) \cite{FirstBECs}, a great deal of experimental and
theoretical efforts has been invested into studies of nonlinear
coherent matter waves. The intrinsic nonlinear nature of BECs,
originating from the interatomic interactions (accounted for by an
effective mean field), together with the techniques available for
the generation of diverse initial configurations in the condensates,
and the tunability of both external trapping potentials and
intrinsic nonlinearity, have made it possible to study an extremely
rich variety of nonlinear macroscopic excitations
\cite{BECBOOK,revnonlin}. In this context, it is relevant to stress
that the effective dimensionality may be chosen as corresponding to
the underlying three-dimensional (3D) geometry, or reduced to nearly
2D (pancake-shaped) and 1D (cigar-shaped) settings, by applying a
strong confinement in the ``undesirable" direction(s)
\cite{BECBOOK}. On the other hand, the $s$-wave scattering length,
which determines the nonlinearity strength in the framework of the
mean-field description of the BEC, may be manipulated by dint of the
magnetically- \cite{mfr}, optically- \cite{ofr-rempe} or
confinement- \cite{cfr} induced Feshbach resonances. In combination
with the use of properly designed trapping (magnetic and/or optical)
potentials, such as optical lattices \cite{OLs}, this technique
paves the way towards a versatile and exceptionally accurate control
over the nonlinear matter waves. Among so generated nonlinear modes,
\textit{matter-wave solitons} have been observed in a series of
famous experiments. In particular, nearly-1D bright and dark
solitons were created in BECs with, respectively,
attractive \cite{Strecker:02,Khaykovich:02,bright3} and repulsive
\cite{DS1,DS2,hamburg,hambcol,kip,technion,peter} interatomic interactions. Bright
solitons of the gap type have also been created in the quasi-1D repulsive
BEC \cite{Eiermann:04} loaded into an optical lattice.

The description of the dynamics of matter waves is based on the
Gross-Pitaevskii equation (GPE) \cite{BECBOOK}, which bears significant
similarities to the nonlinear Schr\"{o}dinger equation (NLSE) governing the
transmission of light signals in nonlinear optical media. Accordingly, the
matter-wave solitons are counterparts of optical solitons, that have been
studied in detail, in temporal, spatial \cite{Kivshar} and spatio-temporal
\cite{review} settings alike.

In this work, we focus on dark solitons (see the reviews
\cite{Kivshar-LutherDavies,djf} for optical and matter-wave dark solitons, respectively),
in connection to higher-dimensional geometry. In that case, an important issue is the
stability of dark matter waves, which was first
analyzed in the framework of the (2+1)-dimensional NLSE.
In particular, the stability of the dark-soliton stripe (DSS) was
studied in Ref.~\cite{kuzne} (see also Refs.~\cite{kuz2,kuz3}), where it was
shown that the DSS is prone to the transverse modulational instability (MI)
(alias ``snaking instability'')
against transverse long-wavelength perturbations. Experimental
\cite{DSinstability1,DSinstability2} and theoretical
\cite{Kivshar-LutherDavies,pelkiv} studies of this instability in the
context of nonlinear optics have revealed that it may lead to
splitting of the DSS into a chain of vortices with alternating
topological charges (vortex--anti-vortex pairs). In particular, it
was found that a quiescent (``black" ) DSS is vulnerable to
transverse ``snaking" deformations
\cite{Kivshar-LutherDavies,pelkiv}, causing the splitting into
vortex pairs. Unstable moving (``gray") DSSs do not split into
vortices, but rather emit radiation in the form of sound waves. In
the BEC context, the transverse MI and splitting of DSSs into vortex
rings was first observed in the experiment with a two-component BEC
composed of $^{87}$Rb atoms in two different hyperfine states
\cite{BPA:01}. In that work, a DSS
was created in
one component, and the snaking instability caused it to decay into
vortex rings (in a quasi-spherical geometry), in accordance with the
theoretical predictions \cite{Feder:00}.

Several works proposed possibilities of suppressing the snaking instability
of DSSs. In particular, in the context of nonlinear optics it is known \cite%
{Kivshar-LutherDavies} that this instability can be avoided in finite-size
holding optical beams \cite{spatial1}. Similarly, in the BEC context, it was
shown \cite{PanosAvoidingRedCatastrophe} (see also Ref.~\cite%
{BrandReinhardt:02}) that the snaking instability may be suppressed in
sufficiently strong traps. On the other hand, it has been shown that 2D DSS
can be stabilized by nonlocal nonlinearities \cite{Kivshar:08}. Furthermore,
3D dark solitons \cite{Nath:08} may be stabilized in dipolar condensates,
which are characterized by long-range dipole-dipole interactions (see a
recent review of this topic in Ref. \cite{pfau}). In that case, the
suppression of the DSS instability is provided by the nonlocal character of
the respective mean-field model (see also Ref.~\cite{strillo} for similar
results in the context of optics). A more complex dark-soliton
configuration, which is not subject to the MI, was reported in
Refs.~\cite{crosses,crosses2}
in the context of the two-component BEC: it features a ``cross" formed by the intersection
of two rectilinear domain walls, with the wave functions of the same
species filling each pair of opposite quadrants, with a $\pi $ phase
difference between them. In this way, a quasi-dark-soliton
configuration is formed, which is stable for long times (even in the
rotating trap) in a large parametric region.

In the present work, we propose and analyze an experimentally relevant
scheme for suppressing and controlling the transverse instability of DSSs in
BECs and optics, based on the use of a repulsive quasi-1D potential barrier,
which can be generated by a blue-detuned laser beam in BECs (see, e.g., Ref.~%
\cite{odt}), or by a slab with a lower value of the refractive index
embedded into the self-defocusing bulk waveguide. This barrier, which repels
the atoms in the condensate or the light in the waveguide, creates a channel
where a DSS can be naturally trapped. Analyzing the transverse MI of the DSS
in this setting, we show that the channel can suppress the snaking
instability. In fact, results produced by our linear-stability analysis and
corroborated by direct simulations indicate that the wave number of the most
unstable perturbation mode is shifted due to the presence of the potential
barrier. The control of the most unstable mode is useful not only for the
complete stabilization of the DSS, but also for controlling the number of
vortex pairs generated by the snaking instability. This control mechanism
can be adjusted by appropriately tuning the height and transverse width of
the potential barrier.

The paper is organized as follows. In Sec.~\ref{SEC:pert}, we introduce the
model and produce an approximate solution for the DSS, in the presence of
the potential barrier. Using the variational approximation, we obtain a
solution for the DSS and analytically predict the critical wavenumber for
the MI modes, by means of the linear-stability analysis (details for the
latter are described in the appendix). In Sec.~\ref{SEC:NUM}, we numerically
study the transverse MI of the DSS, varying the height and width of the
stabilizing potential barrier. We find that, in the uniform space (i.e., in
the absence of an external harmonic
trapping potential) and with a
sufficiently broad repulsive potential barrier, there exists a critical
value of its strength (height), above which DSSs get \emph{completely
stabilized} against the MI. In the case when the instability is not
completely suppressed, we explore the possibility of controlling the density
of the nucleated vortex pairs for an infinitely long DSS, by tuning the most
unstable mode, via the parameters of the barrier. We then consider the more
realistic model of a harmonically confined BEC (which may be relevant to
optics too, representing a rod-shaped bulk waveguide), and perform numerical
simulations which demonstrate the degree of the control over the number of
the nucleated vortex pairs. In Sec.~\ref{SEC:CONCLU}, we summarize the
findings and point out directions for future work.

\section{The model and its analytical consideration \label{SEC:pert}}

\subsection{The fundamental equation and dark-soliton solution}

We start by considering the following (2+1)-dimensional GPE/NLSE in the
usual scaled form, with the repulsive nonlinearity:
\begin{equation}
i\frac{\partial u}{\partial t}=-\frac{1}{2}\left( \frac{\partial ^{2}u}{%
\partial x^{2}}+\frac{\partial ^{2}u}{\partial y^{2}}\right)
+|u|^{2}u+V(x,y)u.  \label{NLS}
\end{equation}%
In the BEC context, this equation governs the evolution of the macroscopic
wave function $u(x,y,t)$ of the pancake-shaped condensate in the $(x,y)$
plane \cite{BECBOOK,revnonlin}. The external potential is assumed to have
the following form:
\begin{equation}
V(x,y)=V_{\mathrm{HT}}(x,y)+V_{\mathrm{LS}}(x),  \label{POT}
\end{equation}%
that includes an external \textit{harmonic trap} with frequencies $\omega
_{x,y}$,
\begin{equation}
V_{\mathrm{HT}}(x,y)=\frac{1}{2}\left( \omega _{x}^{2}x^{2}+\omega
_{y}^{2}y^{2}\right) ,  \label{MT}
\end{equation}%
and the potential barrier (corresponding to the
far-detuned \textit{laser
sheet} illuminating the BEC) with the Gaussian profile:
\begin{equation}
V_{\mathrm{LS}}(x)=A\,\exp \left[ -x^{2}/\left( 2\sigma ^{2}\right) \right] .
\label{PT}
\end{equation}%
Here $A$ and $\sigma $ measure the height and width of the barrier, with $A>0
$ ($A<0$) corresponding to the blue- (red-) detuned laser beam, which repels
(attracts) atoms in the condensate (note that $A<0$
corresponds to a potential trough of
depth $|A|$, rather than a barrier).

In terms of nonlinear optics, Eq.~(\ref{NLS}) governs the evolution of the
local amplitude of the electromagnetic wave in the self-defocusing bulk
waveguide, with $t$ replaced by the propagation distance $z$, while $x,y$
are the transverse coordinates, and $V(x,y)$ describes a local modulation of
the refractive index. The barrier potential (\ref{PT}) then corresponds to a
slab with a lower value of the refractive index embedded into the waveguide,
while potential (\ref{MT}) approximates for the global waveguiding structure.

In the absence of the external potential, $V(x,y)=0$, Eq.~(\ref{NLS}) admits
an exact analytical solution for the DSS. Assuming that its nodal plane is
oriented along the $x$-direction, to be aligned with the potential barrier,
when it is switched on, the DSS solution is
\begin{equation}
u(x,y,t)=\sqrt{\mu }\{\mathcal{B}\tanh [\sqrt{\mu }\mathcal{B}(x-vt)]+i%
\mathcal{A}\}e^{-i\mu t}.  \label{gen1}
\end{equation}%
Here, $\sqrt{\mu }\exp (-i\mu t)$ is the stationary background with density $%
\mu $ (which is actually equal to the normalized chemical potential)
supporting the dark soliton. Further, $\sqrt{\mu }\mathcal{B}$ and $v=\sqrt{%
\mu }\mathcal{A}$ represent, respectively, the soliton's depth (or
inverse width) and velocity, with $\mathcal{A}$ and $\mathcal{B}$
subject to the constraint $\mathcal{A}^{2}+\mathcal{B}^{2}=1$. The
DSS with $\mathcal{A}=0$ ($\mathcal{A}\neq 0$) is stationary
(moving), and is usually called a ``black" (``grey") dark soliton.
Below, we first employ the variational approximation to find an
approximate stationary
DSS solution to Eq.~(\ref{NLS}) in the presence of the potential barrier (%
\ref{PT}), and then perform the linear-stability analysis to study the
transverse MI.

\subsection{Dark-soliton stripe steady state in the presence of the
potential barrier.}

In the absence of the harmonic trapping potential (\ref{MT}), but in the
presence of barrier (\ref{PT}), the profile of a stationary DSS solution to
Eq.~(\ref{NLS}) is sought for as
\begin{equation}
u(x,y,t)=U(x)e^{-i\mu t},  \label{SOLUTION}
\end{equation}%
where $\mu$ is the chemical potential (or $-\mu$ is the propagation
constant, in terms of the optical model),
and $U(x)$ is a real function. Substituting this into Eq.~(\ref{NLS}) leads
to an ordinary differential equation,
\begin{equation}
\frac{1}{2} U^{\prime \prime }+\left[ \mu -Ae^{-x^{2}/\left( 2\sigma
^{2}\right) }\right] U-U^{3}=0.  \label{ODE}
\end{equation}

\begin{figure}[tbp]
\begin{center}
\includegraphics[width=8.0cm]{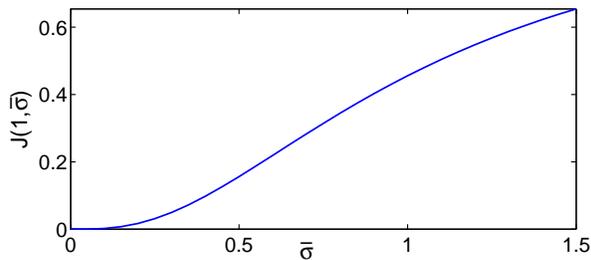}
\end{center}
\par
\vskip-0.4cm
\caption{(Color online) Integral (\protect\ref{EFF_J}) as a function of the
rescaled width $\overline{\protect\sigma }=\protect\beta \protect\sigma $. }
\label{J_vs_sigma}
\end{figure}

As Eq.~(\ref{ODE}) describes both the dark soliton and the uniform
background, one should subtract the contribution of the latter into the
respective Lagrangian, which will be used as the basis of the variational
approximation. To this end, we follow Ref.~\cite{Kivshar:95} (see also Ref.~%
\cite{fr4}), and write the Lagrangian as follows,
\begin{equation}
L=(U^{\prime })^{2}-2\left[ \mu -Ae^{-x^{2}/\left( 2\sigma ^{2}\right) }%
\right] (U^{2}-\mu )+(U^{4}-\mu ^{2}),  \label{VA}
\end{equation}%
where $\mu $ is the amplitude of the uniform background, as defined above.
Then, to approximate the stationary dark-soliton solution of Eq.~(\ref{ODE})
we use the following ansatz [cf.~Eq.~(\ref{gen1})]
\begin{equation}
U(x;A,\sigma )=\sqrt{\mu }\tanh (\beta (A,\sigma )x),  \label{ANS}
\end{equation}%
where the inverse width of the soliton, $\beta $, is a variational
parameter, while the density $\mu $ is the fixed background chemical
potential. Inserting ansatz (\ref{ANS}) into the Lagrangian (\ref{VA}) and
performing the integration over $-\infty <x<+\infty $, we arrive at the
effective (averaged) Lagrangian:
\begin{equation}
L_{\mathrm{eff}}=\frac{4}{3}\mu \left( \beta +\frac{\mu }{\beta }\right)
-2A\mu \int_{-\infty }^{+\infty }e^{-x^{2}/\left( 2\sigma ^{2}\right) }%
\mathrm{sech}^{2}(\beta x)dx.
\end{equation}%
Further, we use the Euler-Lagrange equation, $\partial L_{\mathrm{eff}%
}/\partial \beta =0$, to derive the following implicit equation for the
inverse soliton's width $\beta $:
\begin{equation}
\beta ^{2}=\mu -3A\beta ^{2}J(\beta ,\sigma ),  \label{APP}
\end{equation}%
where
\begin{equation}
J(\beta ,\sigma )\equiv \int_{-\infty }^{+\infty }xe^{-x^{2}/\left( 2\sigma
^{2}\right) }\mathrm{sech}^{2}(\beta x)\tanh (\beta x)dx.  \label{EFF_J}
\end{equation}%
Using the rescaling, $x\rightarrow \beta x$, we conclude that $J(\beta
,\sigma )=J(1,\overline{\sigma })/\beta ^{2},$ where $\overline{\sigma }%
\equiv \beta \sigma $, hence one can rewrite Eq.~(\ref{APP}) as
\begin{equation}
\beta ^{2}=\mu -3AJ(1,\beta {\sigma }).  \label{APP2}
\end{equation}%
Apparently, Eq.~(\ref{APP2}), with $J$ being a function of the single
argument, which is depicted in Fig.~\ref{J_vs_sigma} versus rescaled width $%
\overline{\sigma }$, is simpler than Eq. (\ref{APP}).

\begin{figure}[tbp]
\begin{center}
~\includegraphics[width=8.0cm]{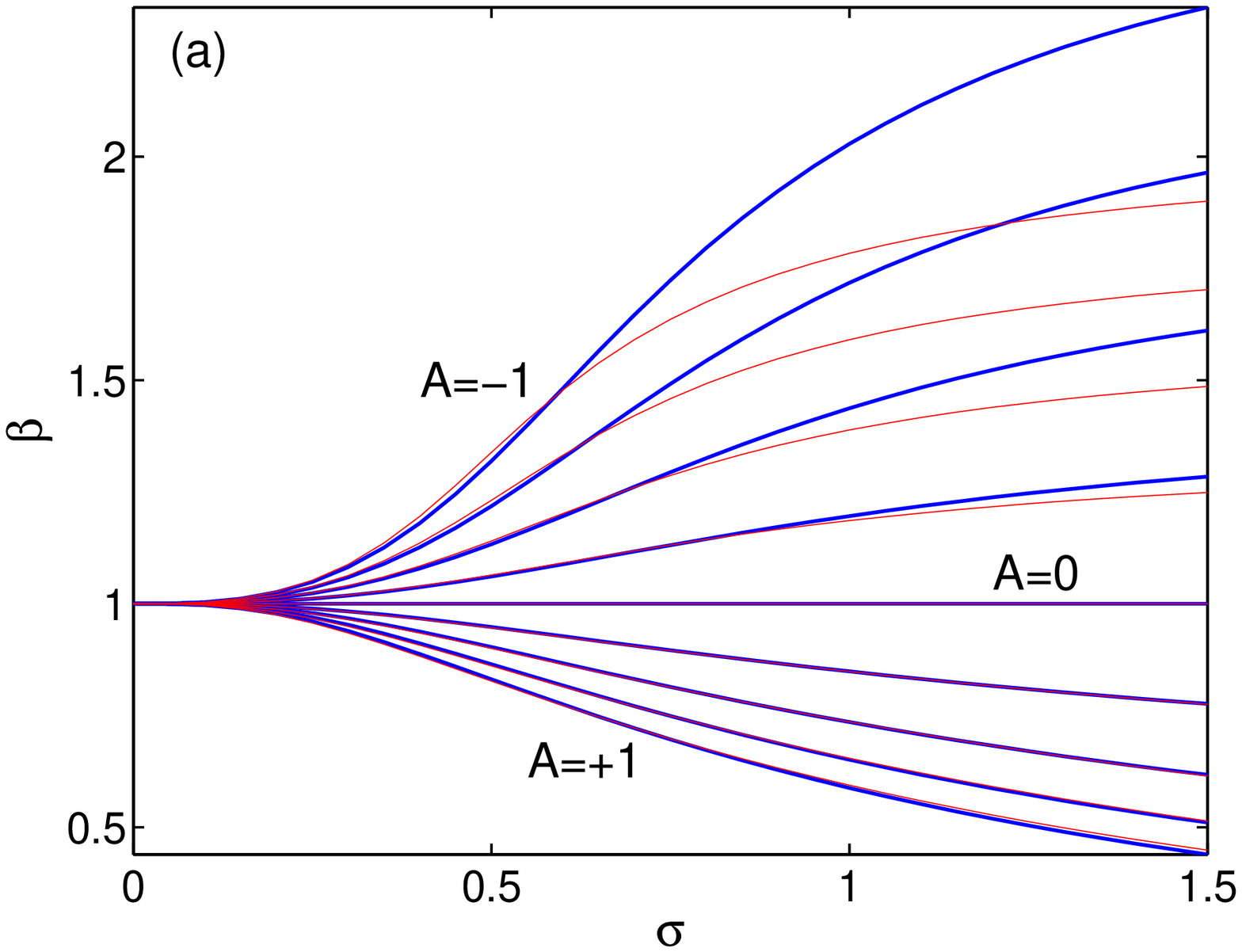}\\[2.0ex]
\includegraphics[width=4.15cm,height=3.0cm]{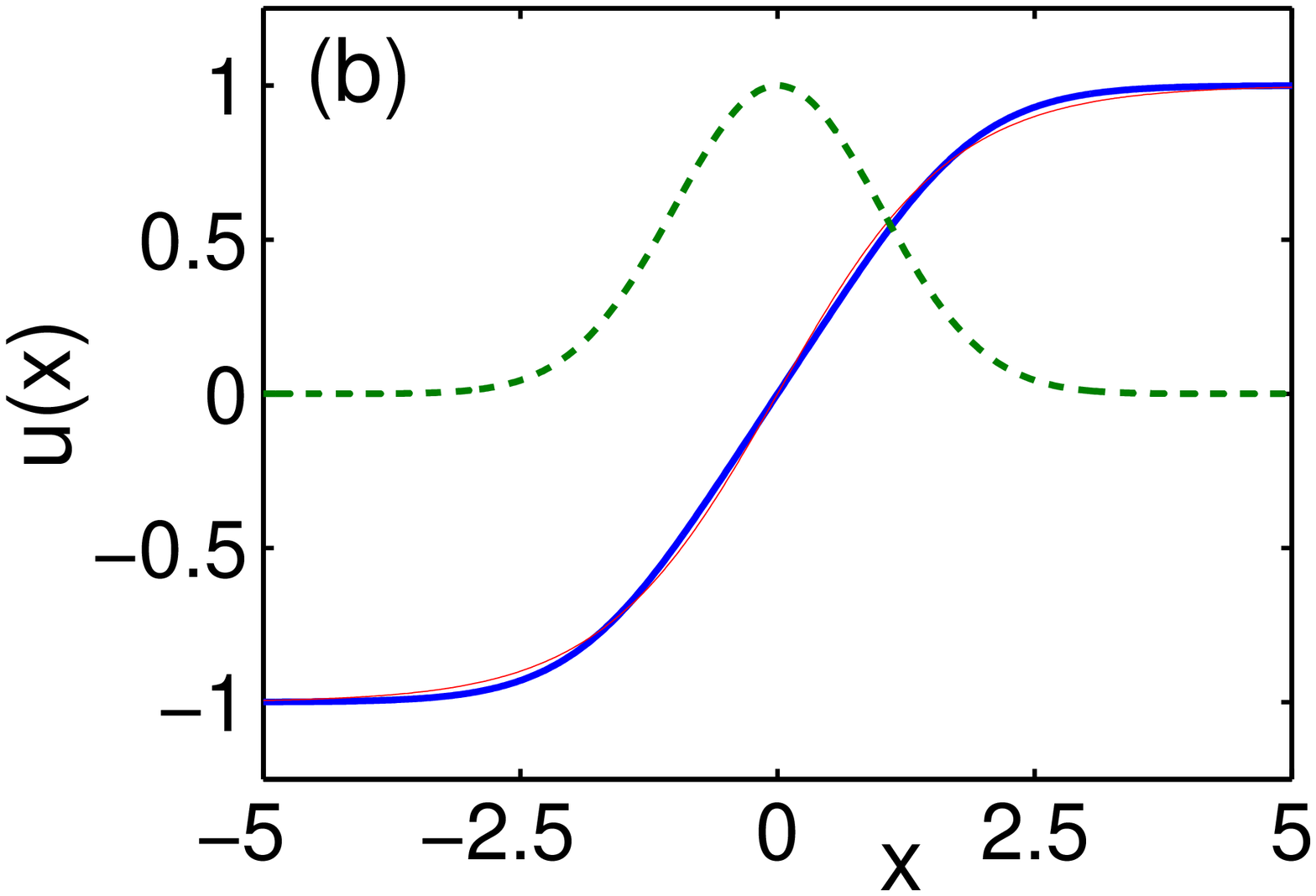} %
\includegraphics[width=3.55cm,height=3.0cm]{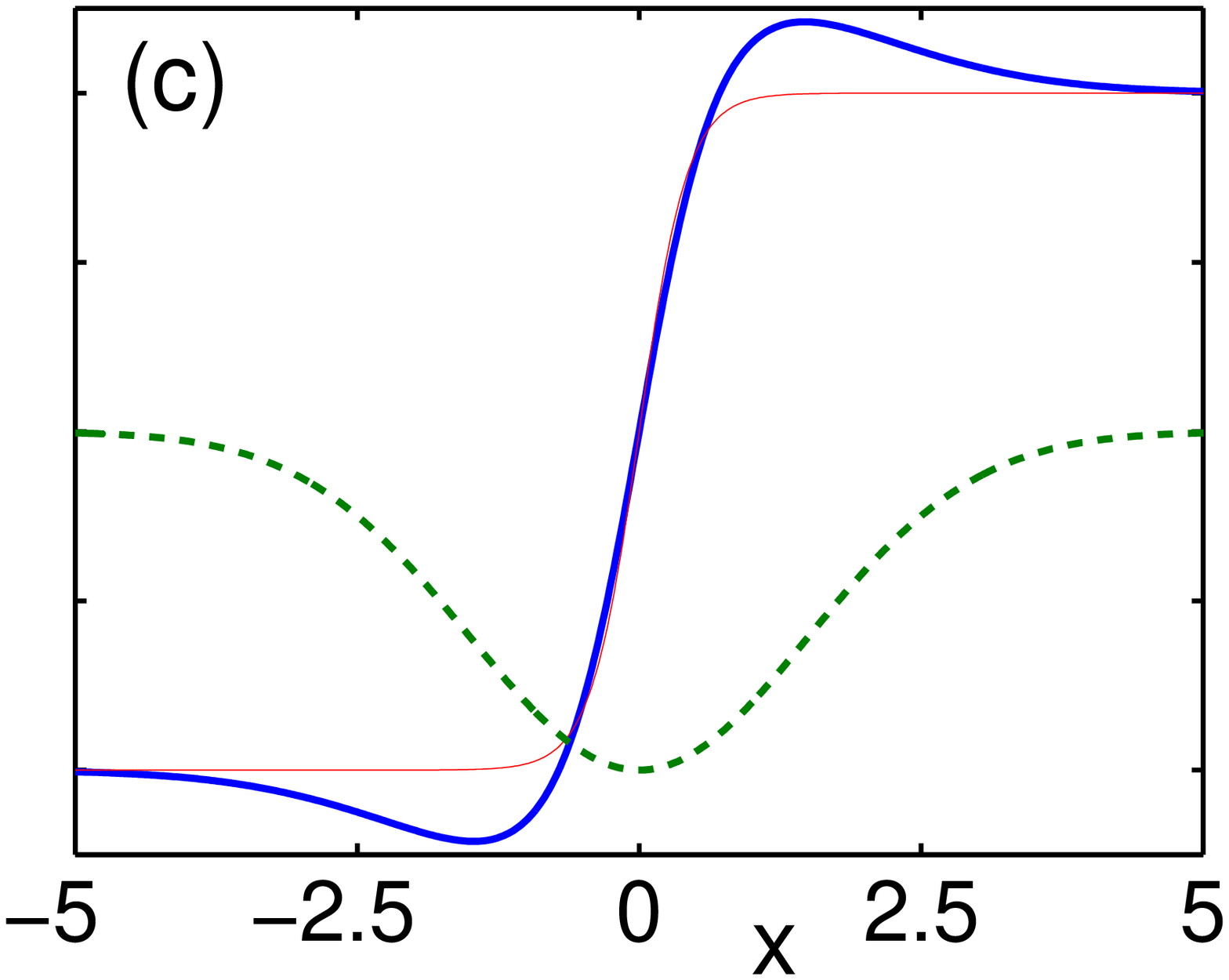}
\end{center}
\par
\vskip-0.4cm
\caption{(Color online) (a) Inverse width $\protect\beta $ of the dark
soliton as a function of width $\protect\sigma $ of the potential barrier,
with height $A$ ranging from $A=-1$ (top curves) to $A=+1$ (bottom curves)
in steps of $0.25$. The thick (blue) curves depict the width of the
numerical solution, obtained by fitting its profile to $\tanh (\protect\beta %
x)$. The thin (red) lines represent the results predicted by the variational
approximation, see Eq.~(\protect\ref{APP2}). Panels (b) and (c) depict,
respectively, two solutions for $(A,\protect\sigma )=(1,1)$ and $(A,\protect%
\sigma )=(-1,1.5)$, the latter case corresponding to the potential trough,
rather than a barrier. The thick (blue) line depicts the numerically found
steady state, while the thin (red) line represents the fitted $\tanh (%
\protect\beta x)$ profile. The transverse profile of the potential barrier
is depicted by the dashed (green) line. The chemical potential is fixed to
be $\protect\mu =1$.}
\label{beta_vs_sigma}
\end{figure}

In Fig.~\ref{beta_vs_sigma} we validate the results of the variational
approximation by comparing the width extracted from the numerically found
steady state, and from the variational equation (\ref{APP2}) [thick (blue)
and thin (red) lines, respectively], for different values of the barrier's
parameters. As seen in the figure, for $A>0$ (i.e., for the repulsive
potential barrier), the variational approximation is very accurate, with the
steady-state solution closely resembling the $\tanh $ profile of ansatz (\ref%
{ANS}) [see panel (b) in the figure]. On the other hand, it is seen in panel
(c) that for $A<0$ (the attractive potential trough, instead of the
barrier) the approximation deteriorates, especially for large widths of the
trough ($\sigma >0.6$, which exceeds the width of the DSS). This observation
is explained by the fact that the steady-state profile develops a localized
hump induced by the attractive trough (see also Ref.~\cite{fr1}), which is
not captured by our ansatz.

\subsection{The transverse modulational instability of the dark-soliton stripe
in the presence of the potential barrier.}

We now consider the solution of Eq.~(\ref{NLS}) in the form of the DSS in
the presence of the potential barrier. In this case, the steady state DSS
may be approximated by
\begin{equation}
u(x,y,t)=u_{0}(x,y)\,e^{-i\mu t}=\sqrt{\mu }\tanh (\beta x)\,e^{-i\mu t},
\label{dark}
\end{equation}%
where $\beta $ is determined by Eq.~(\ref{APP2}). To analyze the MI of this
DSS, we follow the approach of Ref.~\cite{kuzne} and consider small
transverse perturbations, see details in the appendix. The result of the
analysis presented in the appendix is that the critical wavenumber which
defines the width of the instability band for solution (\ref{SOLUTION}) is
\begin{equation}
k_{\mathrm{cr}}=\beta ,  \label{crwave0}
\end{equation}%
where $\beta $ is determined by Eq.~(\ref{APP2}). That is, the DSS (\ref%
{dark}) 
is unstable against transverse perturbations with wavenumbers $k<k_{\mathrm{%
cr}}=\beta $. When this instability sets in, the DSS undergoes a snake-like
transverse deformation, which eventually breaks it into vortex-antivortex
pairs. This outcome of the evolution can be observed in Fig.~\ref%
{stripes_Om0.ps}.

\begin{figure}[tbp]
\begin{center}
\includegraphics[width=8.5cm]{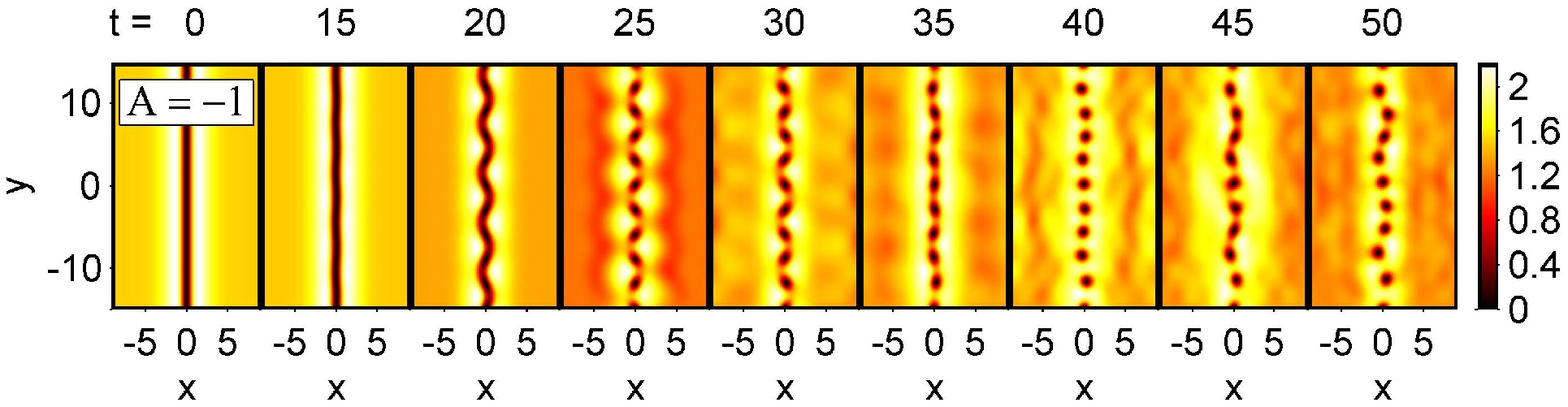} \\[2.0ex%
]
\includegraphics[width=8.5cm]{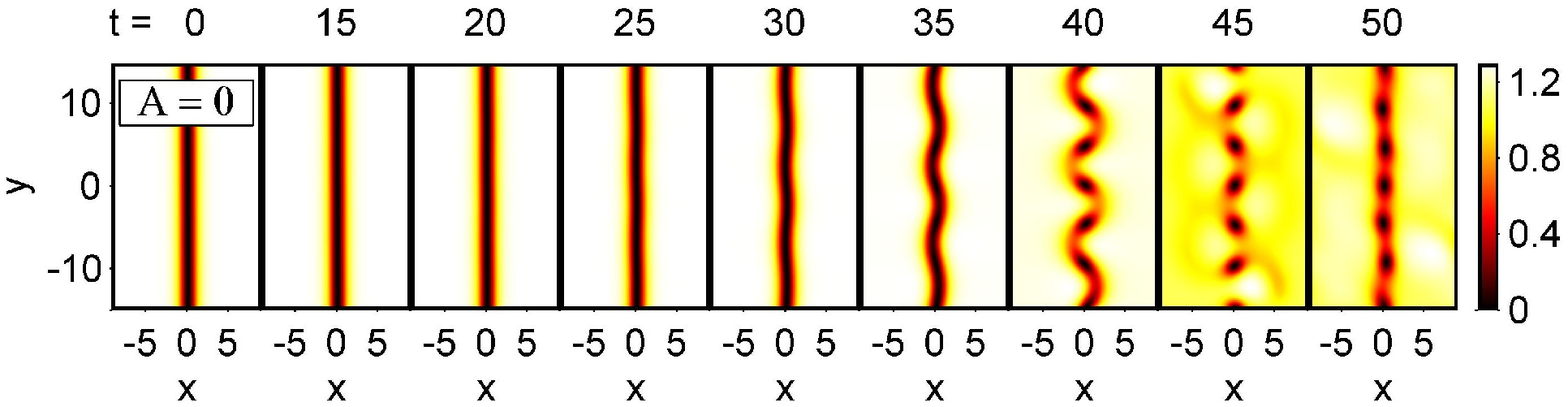} \\[2.0ex]
\includegraphics[width=8.5cm]{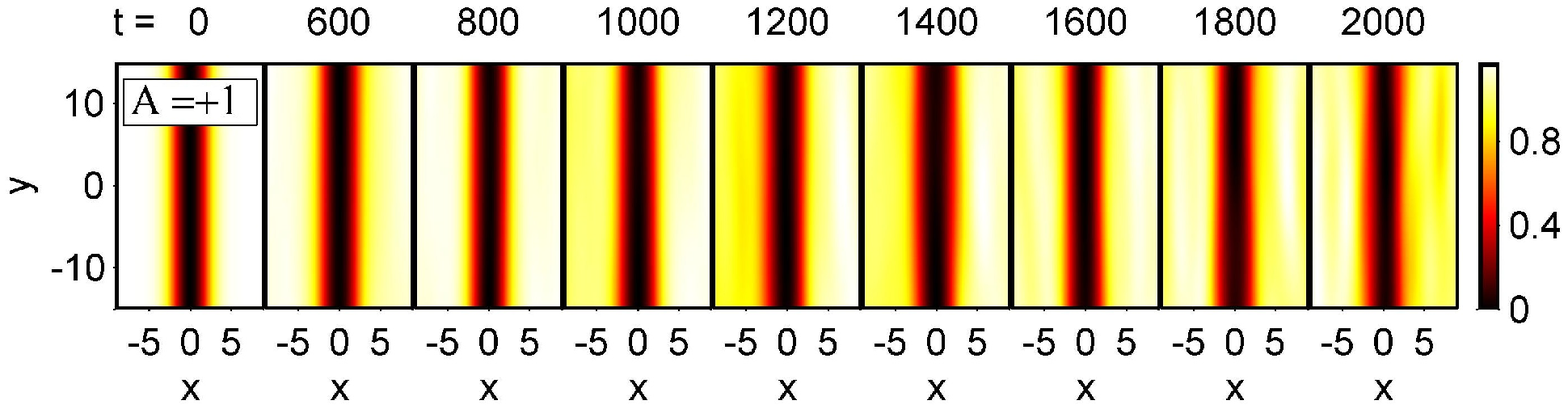}
\end{center}
\par
\vskip-0.4cm
\caption{(Color online) The dynamics induced by the transverse modulational
instability of a dark-soliton stripe in the absence of the external harmonic
trap. Shown are different density snapshots at indicated moments of time.
The initial condition corresponds to a stationary stripe with an added
random perturbation of relative size $10^{-5}$. The top panel: the case of
the attractive potential trough, with $A=-1$ and $\protect\sigma =1.5$. The
middle panel: no potential barrier or trough ($A=0$). The bottom panel: the
case of the repulsive potential barrier, with $A=1$ and $\protect\sigma =1.5$%
. Note the faster destabilization and nucleation of a larger number of
vortices (see dark circular spots for later times) in the case of the
attractive potential trough (the top panel), in comparison to the case when
the trough is absent (the middle panel), and the \emph{complete stabilization%
} of the dark-soliton stripe by the repulsive potential barrier in the
bottom panel. The simulations in this figure were carried out in the domain
of $(x,y)\in \lbrack -15,15]\times \lbrack -15,15]$. }
\label{stripes_Om0.ps}
\end{figure}

From Eq.~(\ref{crwave0}) we can derive the critical number of vortices in
the chain produced by the splitting of the DSS (\ref{dark}):
\begin{equation}
N_{\mathrm{cr}}=\beta L/\pi ,  \label{vortice}
\end{equation}%
where $L$ is the length of the DSS. Since $\beta $ is implicitly determined
by height $A$ and width $\sigma $ of the potential barrier (\ref{PT}), as
per Eq.~(\ref{APP2}), it is possible to control the critical wavenumber by
appropriately adjusting the barrier's parameters. This, in turn, offers a
control of the upper bound of the vortex pairs that are nucleated by the
transverse instability of the DSS. In the next section we show that it is
indeed possible to control the number of the nucleated vortex pairs, which
we compare with the analytical prediction.

\begin{figure}[tbp]
\begin{center}
\includegraphics[width=8.5cm]{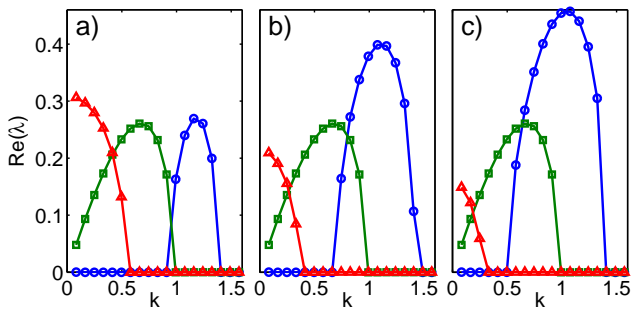}
\end{center}
\par
\vskip-0.4cm
\caption{(Color online) The growth rate of the transverse modulational
instability of dark-soliton stripes at different values of the height and
width of the potential barrier, $A$ and $\protect\sigma $. The panels
correspond to the increasing width: (a) $\protect\sigma =0.5$, (b) $\protect%
\sigma =1$, and (c) $\protect\sigma =1.5$, as indicated by the dashed lines
in Fig.~\protect\ref{plot_allwaveunst.ps}. Each panel displays the real part
of the eigenvalue, for the potential trough/barrier with heights $A=-1$
[(blue) circles], $A=0$ [(green) squares], and $A=1$ [(red) triangles]. The
computations corresponding to this figure were carried out in the domain of $%
(x,y)\in \lbrack -50,50]\times \lbrack -50,50]$. }
\label{plot_allwaveunst_cuts.ps}
\end{figure}

\section{Numerical results\label{SEC:NUM}}

\subsection{Transverse modulational instability of the dark-soliton stripe
in the absence of the harmonic trap \label{subSEC:NOTRAP}}

The main subject of this work is the control of the transverse MI of the DSS
by means of the potential barrier. Typical examples of the snaking
instability are depicted in Fig.~\ref{stripes_Om0.ps}, in the absence of the
external harmonic trap [$\omega _{x,y}=0$ in Eq.~(\ref{MT})]. The figure
depicts the evolution of the density $|u(x,y,t)|^{2}$ at different times,
starting from a stationary DSS with a small initial random perturbation
added to it. The middle panel in the figure depicts the case of a quiescent
(black) DSS ($\beta =1$), with no potential barrier $A=0$, where the
instability manifests itself with wavenumber $k\approx 2\pi /10=0.628$
(wavelength $\simeq 10$), which is smaller than the predicted critical
wavenumber, $k_{\mathrm{cr}}=\beta =1$. In fact, this unstable wavenumber is
the most unstable one, $k_{\mathrm{max}}$, for this configuration.

In order to quantify the instability and to
compare it to the analytical results of the previous section, we compute
the stability spectra of the stationary solutions $u_{0}(x,y)$ through the
standard Bogoliubov-de Gennes (BdG) analysis. This analysis involves
numerically solving the linear eigenvalue BdG problem, which stems from the
linearization of the GPE (\ref{NLS}) around the steady state solution $u_{0}$
by using ansatz $u(x,y,t)=\{u_{0}(x,y)+\left[ a(x,y)e^{\lambda t}+b^{\star
}(x,y)e^{\lambda ^{\star }t}\right]\} e^{-i \mu t} $. The solution of this BdG eigenvalue
problem yields the eigenfunctions $\{a(x,y),b(x,y)\}$ and the eigenvalues $\lambda $.
Note that, due to the Hamiltonian nature of the system, if $\lambda $ is
an eigenvalue of the BdG spectrum, so are also $-\lambda $, $\lambda^{\ast }$
and $-\lambda ^{\ast }$. Notice that the linear stability condition amounts to
$\mathrm{Re}(\lambda )=0$, i.e., all eigenvalues must be imaginary. In fact,
in connection to the stability analysis described in the previous section,
the BdG eigenvalue $\lambda $ corresponds to eigenfrequency $\Omega $
through relation $\Omega =i\lambda $. Therefore, an unstable eigenvalue
corresponding to $\mathrm{Re}(\lambda )>0$ corresponds, in turn, to an
eigenfrequency with $\mathrm{Im}(\Omega )>0$.

The MI spectrum for the DSS in the case of no
potential barrier ($A=0$) is depicted by the (green) squares in all panels
in Fig.~\ref{plot_allwaveunst_cuts.ps}. The most unstable wavenumber, $k_{%
\mathrm{max}}$, corresponds to the location of the maxima of $\mathrm{Re}%
(\lambda )>0$. Note that, as predicted by the linear
stability analysis, all modes with $k>k_{\mathrm{cr}}=\beta =1$ are stable.
It is also interesting to note that for strong enough attractive ($A<0$)
potential barriers the whole spectrum is shifted to the right, so that small
wavenumbers become stable [see (blue) circles in Fig.~\ref%
{plot_allwaveunst_cuts.ps}]. Furthermore, it is important to note that the
results presented in Fig.~\ref{plot_allwaveunst_cuts.ps} were obtained in a
large box, $(x,y)\in \lbrack -50,50]\times \lbrack -50,50]$, in order to
safely capture small wavenumbers. In contrast, the simulations depicted in
Fig.~\ref{stripes_Om0.ps} were performed in a smaller domain box, $(x,y)\in
\lbrack -15,15]\times \lbrack -15,15]$ so that the case for $A=1$ (the
bottom row in the figure) is rendered stable since the corresponding
unstable wavenumbers do not fit into the integration box. This effect,
discussed further below, becomes important when considering trapped
condensates that inherently possess a finite size.

\begin{figure}[tbp]
\begin{center}
\includegraphics[width=8.5cm]{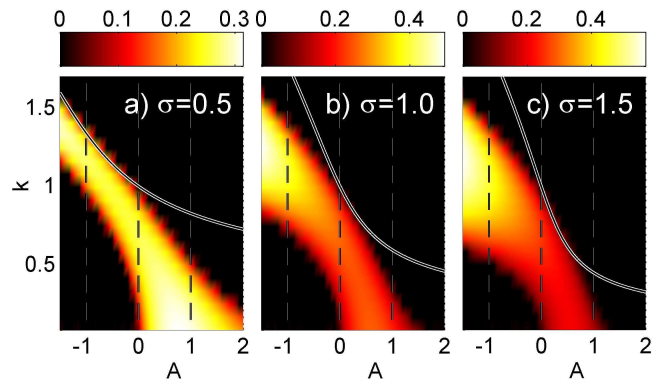}
\end{center}
\par
\vskip-0.4cm
\caption{(Color online) Regions of the transverse modulational instability
of the dark-soliton stripe for different parameters of the potential
barrier. Depicted is the real part of the eigenvalue (the lighter color
represents the stronger instability, while black corresponds to the
stability) as a function of wavenumber $k$ and the barrier's height $A$.
Different panels correspond to increasing width of the barrier: (a) $\protect%
\sigma =0.5$, (b) $\protect\sigma =1$, and (c) $\protect\sigma =1.5$. The
black and white solid lines depict the prediction of the critical wavenumber
as per Eq.~(\protect\ref{crwave0}). Vertical dashed lines represent the cuts
depicted in Fig.~\protect\ref{plot_allwaveunst_cuts.ps}. }
\label{plot_allwaveunst.ps}
\end{figure}

We now turn to the effect of the potential barrier on the stability (and
dynamics) of the DSS. The top panel in Fig.~\ref{stripes_Om0.ps} depicts the
snaking instability corresponding to the attractive potential trough with
strength (depth) $A=-1$. As seen in the figure, the attractive trough
naturally enhances the instability, in comparison to the free-standing DSS,
in two ways: (i) the MI sets in earlier, and (ii) the most unstable
wavenumber is right-shifted ($k_{\mathrm{max}}\approx 2\pi /6=1.0472$). The
corresponding instability spectrum for this case is depicted by the (blue)
circles in Fig.~\ref{plot_allwaveunst_cuts.ps}(c). In fact, comparing the
(green) squares and the (blue) dots in this panel, it is evident that the
net effect of the attractive potential trough is the right-shift of the
instability spectrum. Note that the spectrum is located to the left of the
threshold, $k_{\mathrm{cr}}=\beta \approx 1.78$, which is predicted by the
linear-stability analysis.

In contrast to the destabilizing effects of the attractive potential, the
repulsive barrier stabilizes the DSS, as might be expected and is shown in
the bottom panel of Fig.~\ref{stripes_Om0.ps}. The spectrum for this case is
depicted by the (red) triangles in Fig.~\ref{plot_allwaveunst_cuts.ps}(c).
Note that, although some perturbation modes are still unstable in this case,
their wavelengths are too large to fit the integration domain used in the
simulations shown in the bottom panel of Fig.~\ref{stripes_Om0.ps} and,
thus, the respective sufficiently strongly repulsive barrier renders the DSS
effectively stable (in a still larger domain, we do observe the development
of the weak transverse MI in this case, which is not shown here). The same
stabilizing effect is also observed for other values of the widths of the
barrier, as it can be clearly seen in all spectra depicted in Fig.~\ref%
{plot_allwaveunst_cuts.ps}. This figure depicts, for three values of width $%
\sigma $, the shift of the MI spectrum for the attractive [(blue) circles],
zero [(green) squares] and repulsive [(red) triangles] channel potentials.
The figure suggests that, as mentioned above, for a sufficiently strong
repulsive potential barrier, only very small wavenumbers may be unstable.
Therefore, if the longitudinal size of the condensate (set by the external
trap) or optical waveguide, into which the DSS is embedded, is not too
large, the instability might be fully suppressed, as the unstable
wavelengths would be too large to comply with the longitudinal size (see
also Ref.~\cite{PanosAvoidingRedCatastrophe} and Sec.~\ref{subSEC:TRAP}
below).

In fact, as we now demonstrate, a sufficiently strong repulsive barrier may
completely suppress all instabilities, including those with very small
wavenumbers. In Fig.~\ref{plot_allwaveunst.ps}, we depict the instability
spectra for three different widths of the barrier widths, as a function of
the sign and strength of the channel potential. The light-colored areas in
the panels correspond to unstable modes, with the shading scale
corresponding to the associated instability growth rate for each mode. As
shown in the figure, for sufficiently wide potential channels ($\sigma \geq
1.0$, see the two right panels), there is a positive value of the barrier's
height above which \emph{all} wavenumbers are stable (the zero real part of
the eigenvalues corresponds to the black shade in the panels). Therefore, it
is possible to completely suppress the snaking instability and render the
DSS stable with the appropriate choice of the potential barrier.

For example, for the domain used in Figs.~\ref{plot_allwaveunst_cuts.ps}-\ref%
{plot_allwaveunst.ps} and a potential barrier width of $\sigma=1$ ($%
\sigma=1.5$) and height larger or equal to $A=1.65$ ($A=1.35$) the DSS is
completely stable. We also plot in Fig.~\ref{plot_allwaveunst.ps} (see solid
black and white lines) the critical wavenumber (\ref{crwave0}) obtained in
the analytical form in Sec.~\ref{SEC:pert}. This approximate result (valid
for $A=0$) captures the qualitative behavior of the critical wavenumber, but
fails to give its precise location.

Another important feature of the control over the location of the most
unstable mode $k_{\mathrm{max}}$ is that it allows one to precisely
manipulate the number of nucleated vortices per unit length which emerge
from the snaking instability. The instability nucleates one vortex pair per
snaking wavelength \cite{BrandReinhardt:02}. This feature is clearly visible
at later times in Fig.~\ref{stripes_Om0.ps}. For example, the destabilizing
(attractive) channel potential, with $(\sigma ,A)=(1.5,-1)$, enhances the
outcome by producing $N_{v}=10$ vortices in the domain of length $L=30$,
namely the vortex density $\rho _{v}=10/30=0.33...$ . On the other hand, the
free-standing DSS ($A=0$) is responsible for the nucleation of $\rho
_{v}=6/30=0.2$ vortices per unit length. Thus, one can manipulate the
density of the nucleated vortices, selecting it from zero to a maximum value
which is inversely proportional to the healing length of the condensate, in
the case of BEC (recall that two single-charged vortices cannot coexist at
distances smaller than this length).

\begin{figure}[tbp]
\begin{center}
\includegraphics[width=8.5cm]{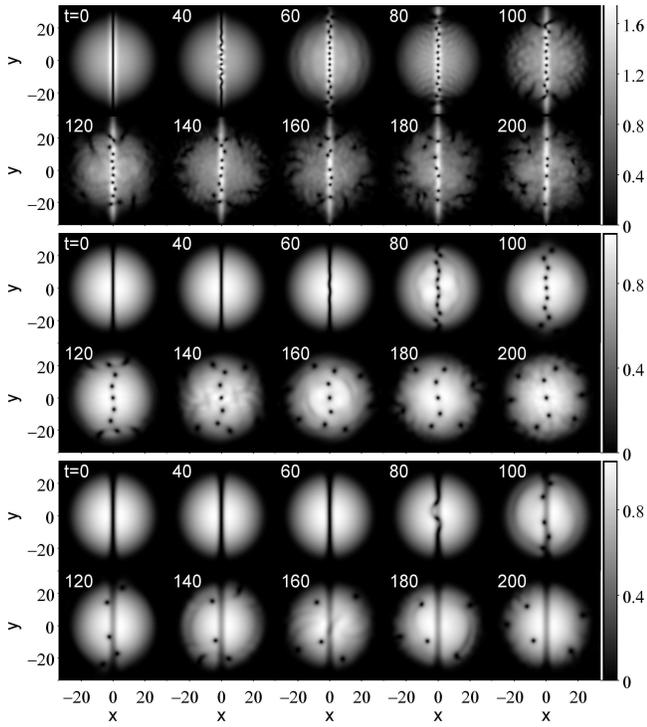}
\end{center}
\par
\vskip-0.4cm
\caption{Dynamics of the transverse modulational instability of a
dark-soliton stripe in the BEC loaded into the harmonic trap of strength $%
\protect\omega =0.05$. Shown are different snapshots of the density at the
times indicated in the figure. The initial condition corresponds to a
stationary stripe to which a random perturbation of relative size $10^{-5}$
was added. The top panel shows the case with the attractive potential trough
of depth $A=-0.8$ and width $\protect\sigma =1.5$. The middle panel: the
situation in the absence of the channel potential ($A=0$). The bottom panel:
the case of the repulsive potential barrier of height $A=0.4$ and width $%
\protect\sigma =1.5$. Note that the number of nucleated vortices can be
controllably varied from about $16$ for $A=-0.8$ (in the top panel) to none
for $A\gtrsim 1$ (not shown here). }
\label{Om0.05.ps}
\end{figure}

\subsection{The modulational instability of dark-soliton stripes in the
harmonic trap \label{subSEC:TRAP}}

We now consider the combined effects of the external harmonic trap and the
channel potential. As mentioned above, this combination can produce an
effective suppression of the snaking instability if the trap's strength is
such that the corresponding Thomas-Fermi (TF) radius of the BEC, $R_{\mathrm{%
TF}}=\sqrt{2\mu }/\omega $ (for $\omega _{x}=\omega _{y}\equiv \omega $), is
smaller than the smallest wavelength of unstable perturbations. In Fig.~\ref%
{Om0.05.ps} we display the development of the DSS' snaking instability and
the concomitant vortex nucleation for the BEC confined in the harmonic trap
of strength $\omega =0.05$. The three panels correspond, from top to bottom,
to channel potentials with width $\sigma =1.5$ and strength $A=-0.8$, $0$,
and $0.4$. It is difficult to precisely measure the number of nucleated
vortices because some of them are nucleated at the rims of the configuration
and are almost invisible. Therefore, we herein focus on measuring the number
of nucleated vortices, $N_{v}$, inside the circle of the TF radius. It is
clear that, for $A=-0.8$, at least $N_{v}=16$ vortices are nucleated, while $%
N_{v}=10$ for $A=0$ (without the channel potential), and only $N_{v}=5$
vortices emerge for $A=0.4$.

\begin{figure}[th]
\begin{center}
\includegraphics[width=8.5cm]{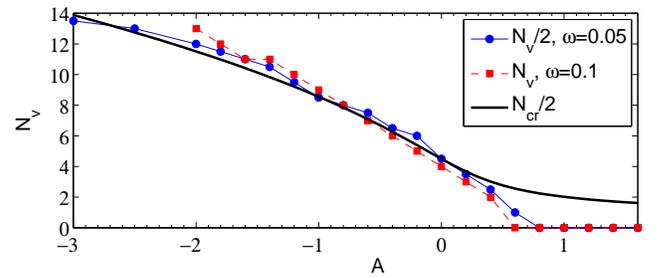}
\end{center}
\par
\vskip-0.4cm
\caption{(Color online) The number of nucleated vortices ($N_{v}$) resulting
from the snaking instability of the dark-soliton stripe in the BEC loaded
into the harmonic trap, versus~the strength ($A$) of the channel
potential.
The number of vortices is measured by direct
count of the number of empty cores present in the density profile after the
snaking instability is fully developed.
The solid line depicts half of the theoretical prediction of
Eq.~(\protect\ref{Ncr}) (see text).
Circles: the trap's strength corresponds to
$\omega =0.05$ and depicted is the half number of the vortices ($N_{v}/2$).
Squares: the trap's strength corresponds to $\protect\omega =0.1$. The width
of potential channel is $\protect\sigma =1.5$ for both data sets.}
\label{num_vortex_vs_A}
\end{figure}

Figure~\ref{num_vortex_vs_A} shows the number of nucleated vortices for two
different values of the trapping frequency. The curve corresponding to $%
\omega =0.05$ has been scaled by factor $1/2$ for a better comparison with
the one corresponding to $\omega =0.1$, since the former has the TF radius
which is twice as large.  In order to estimate
analytically the number of nucleated vortices inside the BEC cloud, we first
establish an upper bound
by considering the number of nucleated vortices in
a DSS of length $L=2R_{\mathrm{TF}}$ with the threshold wavenumber
$k=k_{\mathrm{cr}}$ obtained in Sec.~\ref{SEC:pert}.
Clearly, this estimation is merely
an {\em upper bound} for the number of nucleated vortices because of the following
reasons: (i) since $k_{\mathrm{max}}$ is not readily available from the
theory, we are using $k_{\mathrm{cr}}$ ($k_{\mathrm{max}}<k_{\mathrm{cr}}$),
which corresponds to the instability threshold and not to the wavenumber
with the maximum growth rate; (ii) at the periphery of the TF cloud, it is
difficult to pick the number of vortices through the direct count of the
number of empty cores in the density profiles, and (iii) as it can be
noticed from Fig.~\ref{plot_allwaveunst.ps}, the analytical prediction for
$k_{\mathrm{cr}}$ is always an overestimation of the true wavenumber
threshold. Thus, an upper bound for the number of vortices can be safely
given by
\begin{equation}
N_{\mathrm{cr}}=\frac{\beta L}{\pi }=\frac{2 \beta R_{\mathrm{TF}%
}}{\pi }=\frac{2 \beta \sqrt{2\mu }}{\pi \omega }.  \label{Ncr}
\end{equation}
In connection with item (iii) above, it is worth mentioning that it
is in principle possible to measure the number of vortices by
analyzing the vorticity (curl of the superfluid velocity) of the
condensate. This method reveals (results not shown here) that,
indeed, some vortices created at the periphery of the cloud are
missed by the direct empty-core counting method, based on the
inspection of the density profiles. Nonetheless, we have observed
that these peripheral vortices are quickly pushed out and always
stay at the periphery. Therefore, such boundary vortices do not induce
a considerable change in the long-term BEC\ dynamics. It is also
worth pointing out that counting vortices by the number of empty
cores in the density profile is tantamount to what is often used in
current BEC experiments. Of course, in order to image the vortices
in real experiments, the BEC cloud has to be left to expand and this
could potentially modify the density distribution and ``push"
vortices
inwards or outwards the BEC cloud. However, since the expansion
time is usually short, it is likely that any dynamics during
expansion will be rather slow and, thus, no significant
change in the measurable
number of vortices should be expected.

As concluded above, the
observable number of vortices nucleated by the DSS, $N_{v}$, will
be smaller than the upper bound $N_{\mathrm{cr}}$  defined in
Eq.~(\ref{Ncr}), hence a relation of the type $N_{v}=\alpha N_{\mathrm{cr}}$,
with $\alpha<1$, should be more appropriate.
For the specific value of $\sigma=1.5$ used in Fig.~\ref{num_vortex_vs_A},
we have noticed that the most unstable growth rate $k_{\mathrm{max}}$ is
very close to {\em half} of the critical threshold $k_{\mathrm{cr}}$
for values of the laser intensity $A<1$. Therefore, for this regime,
it is sensible to choose $\alpha =0.5$
as a relevant approximation. In fact, as it can be noticed from
Fig.~\ref{num_vortex_vs_A}, see solid black line, we have found that
the choice of $\alpha =0.5$ gives a very good match to the actual number of
observed vortices in the numerical experiments for a wide range of
values of the potential barrier intensity $A$.
%
It is important, however, to stress that the above estimate, relying on the
assumption that $k_{\mathrm{max}} \approx 0.5 k_{\mathrm{cr}}$,
is only valid for a wide potential barrier with $\sigma=1.5$.
For other regimes, we can only give an upper bound for the number of
nucleated vortices as per Eq.~(\ref{Ncr}).
Furthermore, it is relevant to mention that the prefactor $\alpha$
also depends (results not shown here), for other fixed values of
the laser sheet's width $\sigma$, on the laser strength $A$.

\section{Conclusions\label{SEC:CONCLU}}

In this work, we have proposed and analyzed a technique to control
and even eliminate the
transverse modulational instability, alias the snaking instability, of
dark-soliton stripes in Bose-Einstein condensates (BECs) and bulk
optical waveguides with the self-defocusing nonlinearity. This technique
relies on the use of a
quasi-1D potential barrier, that may be induced by a blue-detuned laser
sheet in the BEC, or by a slab embedded into the optical waveguide. We have
also considered the opposite quasi-1D potential, in the form of an attractive
potential trough (that may be induced by a red-detuning laser beam), which
enhances the snaking instability by making the instability-onset time
shorter and, more importantly, the unstable wavenumbers larger. Therefore,
the attractive trough produces more instability-induced oscillations per
unit length, eventually generating a larger density of vortex-antivortex
pairs. Most importantly, the repulsive barrier is able to suppress the
snaking-instability band, and even \emph{completely stabilize} the
dark-soliton stripe.

We have also developed an analytical approximation for estimating the
largest unstable mode (in the absence of the external harmonic trap). The
analytical results show good agreement with the numerical results obtained
from the Gross-Pitaevskii equation. We also considered the model including
the isotropic harmonic trap. In this case, we monitored the total number of
vortices nucleated by the modulational instability inside the BEC cloud. Assuming that the size of
the cloud is determined by the Thomas-Fermi radius, we were able to
provide an upper bound for the estimate
of the nucleated number of vortices as a function of the
harmonic-trap's strength.

The stabilization mechanism proposed in this work may be naturally extended
in diverse directions. First, it is interesting to consider the
stabilization of quasi-1D solitons by the attractive potential trough in the
model with the self-attraction, and the situation with either sign of the
nonlinearity, if the channel potentials are replaced by a sufficiently
strong periodic optical lattice. Another challenging possibility is to
consider the stabilization of quasi-1D solitons by the cigar-shaped
potential and its periodic optical-lattice counterpart in the 3D geometry.

\section*{Acknowledgments}

R.C.G.~gratefully acknowledges the hospitality of Grupo de F\'{\i}sica No
Lineal (GFNL) of Universidad de Sevilla, and support from NSF-DMS-0806762,
Plan Propio de la University de Sevilla and Grant \# IAC09-I-4669 of Junta
de Andaluc\'{\i}a. M. Ma gratefully acknowledges the hospitality of the
Nonlinear Dynamical Systems (NLDS) group at the San Diego State University,
where this work was carried out with the support from the state scholarship
fund of P. R. China. P.G.K.~gratefully acknowledges support from the
NSF-CAREER program (NSF-DMS-0349023) and from NSF-DMS-0806762.
%
The work of D.J.F.~was partially supported by the Special Account for
Research Grants of the University of Athens. The work of B.A.M.~was, in a
part, supported by the German-Israel Foundation through Grant No. 149/2006,
and by a grant on the topic of ``Nonlinear spatiotemporal photonics in
bundled arrays of waveguides" from the High Council for Scientific and
Technological Cooperation between France and Israel.

\section*{Appendix: The linear stability analysis for the dark-soliton stripe%
}

\label{appA}

In order to analyze the linear stability of the dark-soliton stripe
characterized by Eq.~(\ref{dark}) in the presence of the potential barrier,
we closely follow the approach of Ref.~\cite{kuzne}. We consider a
perturbation of the DSS in the form of
\begin{equation}
\delta u=u_{1}(x)e^{\Lambda },\ \ \delta u^{\ast }=u_{2}(x)e^{\Lambda ^{\ast
}},
\end{equation}%
with $\Lambda =-i\Omega t+iky$, where $\ast $ denotes the complex
conjugation, $\Omega $ is an eigenvalue in the spectrum of Eq.~(\ref{NLS})
linearized around solution~(\ref{dark}), $(u_{1},u_{2}^{\ast })$ is the
corresponding eigenfunction, and $k$ is the corresponding wavenumber along
the $y$ direction.

Substituting the standard ansatz for the perturbation around the soliton
solution (\ref{dark}),
\begin{equation}
u(x,y,t)=(u_{0}+\delta u+\delta u^{\ast })e^{-i\mu t}
\end{equation}%
into Eq.~(\ref{NLS}), and noting the linear independence of $e^{\Lambda }$
and $e^{\Lambda ^{\ast }}$, leads one to the following linear-stability
equations:
\begin{equation}
\left\{
\begin{array}{rcl}
\displaystyle\frac{1}{2}\frac{\partial ^{2}u_{1}}{\partial x^{2}}+(\mu
-V-2|u_{0}|^{2})u_{1} &  &  \\
-u_{0}^{2}u_{2}^{\ast }-\frac{1}{2}k^{2}u_{1}+\Omega u_{1} & = & 0, \\[2ex]
\displaystyle\frac{1}{2}\frac{\partial ^{2}u_{2}^{\ast }}{\partial x^{2}}%
+(\mu -V-2|u_{0}|^{2})u_{2}^{\ast } &  &  \\
-{u_{0}^{\ast }}^{2}u_{1}-\frac{1}{2}k^{2}u_{2}^{\ast }-\Omega u_{2}^{\ast }
& = & 0,%
\end{array}%
\right.   \label{linear2}
\end{equation}%
where the original second equation has been replaced by its
complex-conjugate counterpart. Equation~(\ref{linear2}) can be written in
the matrix form,%
\begin{equation}
\Omega M\varphi -\frac{1}{2}k^{2}I\varphi +L\varphi =0,  \label{matrix}
\end{equation}%
where
\begin{gather}
\varphi \equiv \left(
\begin{array}{c}
u_{1} \\
u_{2}^{\ast } \\
\end{array}%
\right) ,\quad M\equiv \left(
\begin{array}{cc}
1 & 0 \\
0 & -1
\end{array}%
\right) ,\quad I\equiv \left(
\begin{array}{cc}
1 & 0 \\
0 & 1
\end{array}%
\right) ,  \notag \\[0.03in]
L\equiv \frac{1}{2}I\frac{\partial ^{2}}{\partial x^{2}}-\left(
\begin{array}{cc}
2|u_{0}|^{2}-\mu +V & u_{0}^{2} \\
{u_{0}^{\ast }}^{2} & 2|u_{0}|^{2}-\mu +V
\end{array}%
\right) .  \notag
\end{gather}%
Rotating the eigenfunction $\varphi $,
\begin{equation}
\varphi \equiv \left(
\begin{array}{cc}
1 & i \\
1 & -i
\end{array}%
\right) \widetilde{\varphi },
\end{equation}%
yields another representation of Eq.~(\ref{matrix}):
\begin{equation}
\Omega \left(
\begin{array}{cc}
1 & i \\
-1 & i
\end{array}
\right) \widetilde{\varphi }-\frac{1}{2}k^{2}\left(
\begin{array}{cc}
1 & i \\
1 & -i
\end{array}%
\right) \widetilde{\varphi }+L\left(
\begin{array}{cc}
1 & i \\
1 & -i
\end{array}%
\right) \widetilde{\varphi }=0  \label{matrix1}
\end{equation}%
with $\widetilde{\varphi }=\left(
\begin{array}{c}
f_{1} \\
f_{2} \\
\end{array}%
\right) $. Using the matrix $1/2\left(
\begin{array}{cc}
1 & 1 \\
-i & i
\end{array}%
\right) $ to left multiply Eq.~(\ref{matrix1}) yields
\begin{equation}
\Omega M_{i}\widetilde{\varphi }-\frac{1}{2}k^{2}I\widetilde{\varphi }+%
\widetilde{L}\widetilde{\varphi }=0,  \label{matrix2}
\end{equation}%
with $M_{i}=\left(
\begin{array}{cc}
0 & i \\
-i & 0
\end{array}%
\right) $ and $\displaystyle\widetilde{L}=\frac{1}{2}I\frac{\partial ^{2}}{%
\partial x^{2}}
-\left(
\begin{array}{cc}
\Delta  & i\Delta _{-} \\
i\Delta _{-} & \Delta
\end{array}%
\right) ,
$ where $\Delta \equiv 2|u_{0}|^{2}+\Delta _{+}-\mu +V$ and $\Delta _{\pm
}\equiv ({u_{0}^{\ast }}^{2}\pm u_{0}^{2})/2$. By using Eq.~(\ref{dark}), $%
\widetilde{L}$ becomes
\begin{equation}
\widetilde{L}=\frac{1}{2}I\frac{\partial ^{2}}{\partial x^{2}}+\left(
\begin{array}{cc}
L_{1+} & 0 \\
0 & L_{2+}%
\end{array}%
\right) ,
\end{equation}%
where we define
\begin{equation}
L_{1+}=3\mu \,\mathrm{sech}^{2}(\beta x)-2\mu -V,~L_{2+}=\mu \,\mathrm{sech}%
^{2}(\beta x)-V.
\end{equation}%
As follows from Eqs.~(\ref{matrix}) and (\ref{matrix2}), the main piece of
the information concerning the spectrum of eigenvalues $\Omega $ comes from
properties of the operator $\widetilde{L}$. Obviously, it is a Hermitian
operator, hence its spectrum is purely real. The respective eigenvalue
problem can be written as
\begin{equation}
\widetilde{L}\psi =\frac{1}{2}p^{2}\psi ,\ \ p=p^{\ast }.  \label{linear5}
\end{equation}%
We should note that $p$ corresponds to $k$ in Eq.~(\ref{matrix2}) with $%
\Omega =0$. Our purpose is to find the critical value $k_{\mathrm{cr}}$ of
wavenumber $k$ for the instability of solution~(\ref{dark}). To proceed, we
need to look for a solution to (\ref{linear5}) as
\begin{equation}
\psi =\left(
\begin{array}{c}
f_{1} \\
f_{2}%
\end{array}%
\right) =\left(
\begin{array}{c}
e^{-\beta x}g_{1}(x) \\
e^{-\beta x}g_{2}(x)%
\end{array}%
\right) ,  \label{trial}
\end{equation}%
so that the perturbation functions $f_{1}$ and $f_{2}$ share the same
exponential decay rate near $\pm \infty $ as that of the dark soliton $u_{0}$%
, where $g_{1}$ and $g_{2}$ are constants in the limit of $|x|\rightarrow
\infty $, and the derivatives $dg_{i}/dx$ and $d^{2}g_{i}/d{x}^{2}$ are
bounded in the whole real domain so that
\begin{equation}
\lim_{|x|\rightarrow \infty }\frac{dg_{i}}{dx}=0,
\quad
\lim_{|x|\rightarrow \infty }\frac{d^{2}g_{i}}{d{x}^{2}}=0,
\quad
i=1,2.
\label{condition}
\end{equation}
Substituting Eq.~(\ref{trial}) into Eq.~(\ref{linear5}) leads to the
following equations for $g_{1}(x)$ and $g_{2}(x)$:
\begin{equation}
\left\{
\begin{array}{l}
\displaystyle-\beta \frac{dg_{1}}{dx}+\frac{1}{2}\frac{d^{2}g_{1}}{d{x}^{2}}%
\displaystyle \\[2ex]
\displaystyle+
\left(-\frac{1}{2}{p}^{2}+\frac{1}{2}\beta ^{2}-2\mu +3\mu
\mathrm{sech}^{2}(\beta x)-V\right)g_{1}=0, \\[3ex]
\displaystyle-\beta \frac{dg_{2}}{dx}+\frac{1}{2}\frac{d^{2}g_{2}}{d{x}^{2}}
\\[3.0ex]
\displaystyle+
\left(-\frac{1}{2}{p}^{2}+\frac{1}{2}\beta ^{2}+\mu \mathrm{sech}^{2}(\beta
x)-V\right)g_{2}=0.%
\end{array}%
\right.   \label{linear8}
\end{equation}%
Considering the asymptotic behavior of Eq.~(\ref{linear8}) in the limit $%
|x|\rightarrow \infty $, noting that $V\rightarrow 0$ and $\mathrm{sech}%
^{2}(\beta x)\rightarrow 0$, and taking Eq.~(\ref{condition}) into regard,
we obtain
\begin{equation}
\left\{
\begin{array}{l}
\displaystyle\left( -\frac{1}{2}{p}^{2}+\frac{1}{2}\beta ^{2}-2\mu \right)
\tilde{g}_{1}=0, \\[2ex]
\displaystyle\left( -\frac{1}{2}{p}^{2}+\frac{1}{2}\beta ^{2}\right) \tilde{g%
}_{2}=0,%
\end{array}%
\right.   \label{linear9}
\end{equation}%
where $\tilde{g}_{1}\equiv g_{1}(|\pm \infty |)$ and $\tilde{g}_{2}\equiv
g_{2}(|\pm \infty |)$ are constants. Equation (\ref{linear9}) has,
therefore, two cases to discuss.

\noindent \textbf{Case 1:} If $\tilde{g}_{2}=0$, then, from the first
equation in Eq.~(\ref{linear9}), we have $p^{2}=\beta ^{2}-4\mu .$However,
using Eq.~(\ref{APP}) and Eq.~(\ref{EFF_J}), this result yields a sign
contradiction for small $|A|$. Thus, this case does not yield any sensible
solution.

\noindent \textbf{Case 2:} If $\tilde{g}_{1}=0$, then, from the second
equation in Eq.~(\ref{linear9}), we have $p^{2}=\beta ^{2}$. Therefore, the
critical wave number for the instability region of the solution Eq.~(\ref%
{dark}) is given by
\begin{equation}
k_{\mathrm{cr}}=\beta ,  \label{crwave}
\end{equation}%
where $\beta $ is determined by Eq.~(\ref{APP2}). It is interesting to note
that the stability threshold does not depend explicitly on the potential
barrier; this is a consequence of employing the limit of $x\rightarrow \pm
\infty $ in Eq.~(\ref{linear8}), where the tail of the potential decays to
zero. Nonetheless, the effect of the potential barrier is felt, indirectly,
by the DSS through the change of its width $\beta $ determined by Eq.~(\ref%
{APP2}).


\end{document}